\documentclass[iop,apj]{emulateapj}
\usepackage{CJK}
\usepackage{natbib}
\usepackage[dvips]{color}
\usepackage{multirow}
\usepackage{amsmath}
\usepackage{amsfonts}
\usepackage{hyperref}
\usepackage[hyphenbreaks]{breakurl}
\newcommand{\HII}{H\,{\sc ii}}
\newcommand{\HI}{H\,{\sc i}}

\newcommand{\LIR}{$L_{\rm IR}$}

\newcommand{\mum}{$\mu$m}
\newcommand{\herschel}{{\it Herschel}}
\newcommand{\iras}{{\it IRAS}}
\newcommand{\kms}{km\,s$^{-1}$}
\newcommand{\midco}{CO~(6$-$5)}
\begin{document}
%\sloppy
%\begin{CJK*}{UTF8}{gbsn}
%\shorttitle{ALMA \midco\ observation in IC 5179}
\shortauthors{Zhao et al.}
\title{ALMA Maps of Dust and Warm Dense Gas Emission in the Starburst Galaxy IC 5179$^\star$}
\author{Yinghe Zhao\altaffilmark{1,2,3}, Nanyao Lu\altaffilmark{4,5}, Tanio D\'{i}az-Santos\altaffilmark{6}, C. Kevin Xu\altaffilmark{4,5}, Yu Gao\altaffilmark{7}, Vassilis Charmandaris\altaffilmark{8,9}, Paul van der Werf\altaffilmark{10}, Zhi-Yu Zhang\altaffilmark{11,12} Chen Cao\altaffilmark{13,14}
}
\altaffiltext{$\star$}{The National Radio Astronomy Observatory is a facility of the National Science Foundation operated under cooperative agreement by Associated Universities, Inc.}
\altaffiltext{1}{Yunnan Observatories, Chinese Academy of Sciences, Kunming 650011, China; zhaoyinghe@ynao.ac.cn}
\altaffiltext{2}{Key Laboratory for the Structure and Evolution of Celestial Objects, Chinese Academy of Sciences, Kunming 650011, China}
\altaffiltext{3}{Center for Astronomical Mega-Science, CAS, 20A Datun Road, Chaoyang District, Beijing 100012, China}
\altaffiltext{4}{National Astronomical Observatories of China, Chinese Academy of Sciences, Beijing 100012, China}
\altaffiltext{5}{China-Chile Joint Center for Astronomy, Chinese Academy of Sciences, Camino El Observatorio, 1515 Las Condes, Santiago, Chile}
\altaffiltext{6}{N\'{u}cleo de Astronom\'{i}a de la Facultad de Ingenier\'{i}a, Universidad Diego Portales, Av. Ej\'{e}rcito Libertador 441, Santiago, Chile}
\altaffiltext{7}{Purple Mountain Observatory, Chinese Academy of Sciences, Nanjing 210008, China}
\altaffiltext{8}{Department of Physics, University of Crete, GR-71003 Heraklion, Greece}
\altaffiltext{9}{IAASARS, National Observatory of Athens, GR-15236, Penteli, Greece}
\altaffiltext{10}{Leiden Observatory, Leiden University, P.O. Box 9513, 2300 RA Leiden, The Netherlands}
\altaffiltext{11}{Institute for Astronomy, University of Edinburgh, Royal Observatory, Blackford Hill, Edinburgh EH9 3HJ, UK}
\altaffiltext{12}{11 ESO, Karl Schwarzschild Strasse 2, D-85748 Garching, Munich, Germany}
\altaffiltext{13}{School of Space Science and Physics, Shandong University at Weihai, Weihai, Shandong 264209, China}
\altaffiltext{14}{Shandong Provincial Key Laboratory of Optical Astronomy and Solar-Terrestrial Environment, Weihai, Shandong 264209, China}

\date{Received:~ Accepted:~}
\begin{abstract}

We present our high-resolution ($0^{\prime\prime}.15\times0^{\prime\prime}.13$, $\sim$34\,pc) observations of the \midco\ line emission, which probes the warm and dense molecular gas, and the 434\,\mum\ dust continuum emission in the nuclear region of the starburst galaxy IC 5179, conducted with the Atacama Large Millimeter Array (ALMA). The \midco\ emission is spatially distributed in filamentary structures with many dense cores and shows a velocity field that is characteristic of a circum-nuclear rotating gas disk, with 90\% of the rotation speed arising within a radius of $\lesssim150$\,pc. At the scale of our spatial resolution, the \midco\ and dust emission peaks do not always coincide, with their surface brightness ratio varying by a factor of $\sim$10. This result suggests that their excitation mechanisms are likely different, as further evidenced by the Southwest to Northeast spatial gradient of both CO-to-dust continuum ratio and Pa-$\alpha$ equivalent width. Within the nuclear region (radius$\sim$300 pc) and with a resolution of $\sim$34\,pc, the CO line flux (dust flux density) detected in our ALMA observations is $180\pm18$\,Jy\,\kms\ ($71\pm7$\,mJy), which account for 22\% (2.4\%) of the total value measured by \herschel.
\end{abstract}

\section{Introduction}
Luminous infrared galaxies (LIRGs; $L_{\rm IR}$[8$-$1000$\,\mu {\rm m}]>10^{11}$\,$L_\odot$) are a mixture of single galaxies, galaxy pairs, interacting systems and advanced mergers (e.g. Haan et al. 2011, 2013; Petty et al. 2014; Larson et al. 2016; Psychogyios et al. 2016), and their space density exceeds that of optically selected starburst and Seyfert galaxies (AGNs) at comparable bolometric luminosity (Soifer et al. 1987). Compared to less luminous galaxies, LIRGs host a higher fraction of AGN, and exhibit enhanced star formation rates (SFR), usually in the nuclear region (Sanders \& Mirabel 1996). At redshift $z \gtrsim 1$, they dominate the cosmic SFR density (Le Fl\'{o}ch et al. 2005; Caputi et al. 2007; Magnelli et al. 2009, 2011; Gruppioni et al. 2013; though generally high-$z$ LIRGs may have different physical properties compared to their local counterparts (Iono et al. 2009; Menendez-Delmestre et al. 2009). Thus, a detailed  study of local LIRGs is critical to our understanding of the star formation, cosmic evolution of galaxies and AGNs.

LIRGs (including ultra-LIRGs, i.e. $L_{\rm IR}>10^{12}$\,$L_\odot$) in local Universe are all known to be extremely rich in molecular gas, with the ratio of total H$_2$ to \HI\ mass typically $>$1 (Sanders \& Mirabel 1996). While the CO~(1$-$0) line has been widely used to trace the total molecular gas content, star formation occurs in the denser parts of molecular gas clouds as evidenced by the tight correlations between \LIR\ and dense gas tracers such as HCN and CS (e.g., Gao \& Solomon 2004; Wu et al. 2005; Zhang et al. 2014; Privon et al. 2015). Since star formation is expected to heat up the surrounding molecular gas substantially, the resulting warm dense gas can be better traced by the mid-$J$ (e.g. $5\leq J_{\rm upp} \leq 8$) CO line transitions, which have critical densities ($n_{\rm crit}$) of a few times $10^5$\,cm$^{-3}$, and excitation temperatures of $100-200$\,K (Carilli \& Walter 2013).

Indeed, using the Herschel Space Observatory (hereafter \herschel; Pilbratt et al. 2010) SPIRE Fourier Transform Spectrometer (FTS; Griffin et al. 2010) data on a flux-limited subsample of nearby LIRGs selected from the Great Observatories All-Sky LIRG Survey (GOALS; Armus et al. 2009), Lu et al. (2014, 2017) found that, at galactic scales, SFR (as probed by \LIR) correlated much better with the (6$-$5) and (7$-$6) transitions of CO, rather than with the low-$J$ ones ($J\lesssim4$). This result has been  confirmed by Liu et al. (2015) on an expanded sample also including nearby normal galaxies and star-forming regions, in which the authors showed that the best correlations (i.e. smallest dispersion) are between mid-$J$ CO line emission and IR continuum emission (i.e. the star formation law). However, at smaller scale (e.g. $\sim$$0.6-1$ kpc, depending on the specific transition), these correlations start to show larger dispersion and/or deviate from the general relations (e.g. Figure 1 in Liu et al. 2015). Furthermore, our earlier Atacama Large Millimeter Array (ALMA; Wootten \& Thompson 2009) high-resolution imaging of several  LIRGs in \midco\ (Xu et al. 2014, 2015; Zhao et al. 2016b) also showed evidence of the breakdown of star formation law at a physical scale of $\lesssim$100\,pc.

\begin{deluxetable*}{cccccccc}[t]
\centering
\tablecaption{Basic properties of IC 5179 and ALMA Observation Log\label{obslog}}
\tablewidth{0pt}
\tabletypesize{\scriptsize}
\tablehead{
\multicolumn{8}{c}{Basic Properties}
}
\startdata
\multirow{2}{*}{Name} & R.A. (J2000)& Decl. (J2000)& $D_{\rm L}$ & $cz$ & Morph. & Spectral Type &$\log\,L_{
\rm IR}$\\
&(hh:mm:ss)&(dd:mm:ss)& (Mpc)  & (\kms) &&&($L_\odot$)\\
(1)&(2)&(3)&(4)&(5)&(6)&(7)&(8)\\
\hline\noalign{\smallskip}
IC 5179&22:16:09.10& $-$36:50:37.4&51.4&3422&SA(rs)bc&Starburst&11.24\\
\hline\noalign{\smallskip}
\multicolumn{8}{c}{ALMA Observation Log}\\
\hline\noalign{\smallskip}
\multirow{2}{*}{SB}&Date&Time\,(UTC)&Configuration&$N_{\rm ant}$&$l_{\rm max}$&$t_{\rm int}$&$T_{\rm sys}$\\
&(yyyy/mm/dd)&&&&(m)&(min)&(K)\\
(1)&(2)&(3)&(4)&(5)&(6)&(7)&(8)\\
\hline\noalign{\smallskip}
X87b480\_Xb68&2014/07/26&03:10:59$-$04:48:15&C34-5&27&820&39.44&$637-752$\\
Xa25bbf\_X881e&2015/06/07&08:14:06$-$10:08:44&C34-5&35&784&44.15&$701-748$\\
Xb47876\_X1665&2016/06/18&07:53:28$-$08:47:15&C36-4&42&650&17.94&$797-872$
\enddata
\tablecomments{For {\bf basic properties}. Column 1: source name; Columns 2 and 3: right ascension and declination; Column 4: luminosity distance (we use $D_{\rm A}=D_{\rm L}/(1+z)^2$ to get angular size distance); Column 5: Heliocentric velocity from NASA/IPAC extragalactic database (NED); Column 6: morphology classification from NED; Column 7: Nuclear activity classification; Column 8: total infrared luminosity. For {\bf ALMA observation log}. Column 1: schedule-block number; Column 2 and 3: observation date and time; Column 4: configuration; Column 5: number of usable antennae; Column 6: maximum baseline length; Column 7: on-source integration time; Column 8: median system temperature of different SPWs.}
\end{deluxetable*}

 However, the \midco\ line is extremely difficult to observe in nearby galaxies using ground-based facilities, even with ALMA, and thus high-resolution ($\lesssim$kpc) interferometric observations, mostly thanks to ALMA but also the SMA, are available only for a handful of sources: Arp 220 (Matsushita et al. 2009; Rangwala et al. 2015), VV 114 (Sliwa et al. 2013), NGC 34 (Xu et al. 2014), NGC 1068 (Garc\'{i}a-Burillo et al. 2014, 2016), NGC 1614 (Sliwa et al. 2014; Xu et al. 2015; Saito et al. 2017); NGC 253 (Krips et al. 2016), NGC 7130 (Zhao et al. 2016b), and IRAS13120-5453 (Sliwa et al. 2017). Among these observations, about two thirds have physical resolutions of $\lesssim$100 pc, and only the data for NGC 1068 (Garc\'{i}a-Burillo et al. 2014, 2016) have achieved a physical resolution better than 50 pc due to its much closer distance compared to other targets (except for NGC 253).

In order to study the properties of warm molecular gas and cold dust, and their relations to star formation in more details, we have initiated a multi-cycle ALMA program to observe the nuclear regions in representative LIRGs from our FTS sample to map simultaneously the \midco\ line emission (rest-frame frequency $=691.473$\,GHz) and the dust continuum emission at $\sim$434\,\mum, with an ultimate goal of reaching a physical resolution of $\sim$50 pc, i.e. the typical size of a Galactic giant molecular cloud (GMC; Solomon et al. 1979), or less. Our ALMA targets include NGC 34 (Xu et al. 2014) and NGC 1614 (Xu et al. 2015) from the Cycle-0 observations, NGC 7130 (Zhao et al. 2016b) from the Cycle-2 observations, and IC 5179 of which the data presented here are combined from 3 independent observing campaigns across Cycles 2 and 3. The Cycle-0 sources represent advanced mergers with a very warm far-infrared (FIR) color (e.g. the \iras\ 60-to-100 \mum\ ratio $f_{60}/f_{100}$$\sim$1), whereas the Cycle-2 targets have FIR colors ($f_{60}/f_{100}\sim0.5-0.7$) more representative of typical LIRGs, covering both compact nuclear core and circumnuclear disk configurations visible in the high-resolution Pa-$\alpha$ images of Alonso-Herrero et al. (2002; 2006), and whether there exists a significant AGN contribution to the ionized gas heating based on the [Ne\,{\sc v}] observation of Petric et al. (2011). 

Our previous studies have illustrated the different morphologies in the \midco\ emission within the nuclear regions of the three LIRGs: a rotating disk in NGC 34 (Xu et al. 2014), a rotating ring in NGC 1614 (Xu et al. 2015), and co-rotating clouds in NGC 7130 (Zhao et al. 2016b). Moreover, the \midco\ in NGC 7130 shows a unique double-lobed structure, well consistent with the dust lanes in the optical/near-infrared images. Furthermore, there seems to be a trend in the sense that the nuclear region (with a diameter of about 500 pc) contains a smaller fraction of the dense gas emission as the FIR color becomes cooler. 

Whereas NGC 7130 is a peculiar spiral accompanied by two dwarf galaxies, and hosts an AGN, IC 5179 is an isolated, pure starburst galaxy (Veillux et al. 1995) with a morphology type of SA(rs)bc (de Vaucouleurs et al. 1991). The star formation activities in IC 5179 are widespread as demonstrated by the H$\alpha$ emission (Lehnert \& Heckman 1995), and numerous \HII\ regions are distributed in the spiral arms (Alonso-Herrero et al. 2006). Its inclination angle of $\sim$63$^\circ$, calculated by using the axis ratio from Lehnert \& Heckman (1995), indicates that we nearly have an edge-on view of this source. At its distance (angular size distance) of 50.2\,Mpc, 1\arcsec\ corresponds to 244\,pc. As shown in Armus et al. (2009), IC 5179 has $L_{\rm IR}=10^{11.24}\,L_\odot$, and $f_{60}/f_{100}\sim 0.52$, which make it the least luminous and coldest target in our ALMA sample. The nucleus is the most obscured region in this galaxy (Piqueras L\'{o}pez et al. 2013), with annular mean $A_V$ (derived using the Br-$\gamma$/Br-$\delta$ ratio) in the range of $\sim$5-10 mag, consistent with that obtained from the near-IR (NIR) colors (D\'{i}az-Santos et al. 2008).

 Previous single-dish observations of low-$J$ $^{12}$CO and $^{13}$CO transitions (Mirabel et al. 1990; Garay et al. 1993) in IC 5179 show that these lines have multiple peaks. The $^{12}$CO (1$-$0) to $^{13}$CO (1$-$0) line ratio is $\sim$9, consistent with the value for normal galaxies (Sage \& Isbell 1991) but much smaller than that for starburst mergers (Taniguchi \& Ohyama 1998). The CO spectral line energy distribution (SLED) of IC 5179 observed by \herschel\ peaks at $J=4$ (Lu et al. 2017), whereas the CO (4$-$3) to (1$-$0) line ratio is about 0.3 based on the \herschel\ and SEST 15-m data (the beams at these frequencies are roughly the same, i.e., $\sim$45\arcsec; Mirabel et al. 1990; Garay et al. 1993; Chini et al. 1996; Albrecht et al. 2007). Therefore, the CO SLED of IC 5719 is closer to those of high-$z$ submillimeter galaxies (e.g., Casey et al. 2014) compared to most local (U)LIRGs which have  significantly warmer CO SLEDs.

The remainder of this paper is organized as follows. We report our observations and data reduction in Section 2,  present the results and discussion in Section 3, and briefly summarize the main conclusions in the last section. Throughout the paper, we adopt a Hubble constant of $H_0 = 70$\,\kms\,Mpc$^{-1}$, $\Omega_{\rm M} = 0.28$, and $\Omega_{\Lambda}= 0.72$, which are based on the five-year {\it WMAP} results (Hinshaw et al. 2009), and are the same as those used by the GOALS project (Armus et al. 2009).

\section{Observations and Data Reduction}
The observations of the central region of IC 5179 in \midco\ were conducted using the Band 9 receivers of ALMA in the time division mode with a velocity resolution of $\sim$6.8\,\kms. The four basebands (i.e., “Spectral Windows”; SPWs 0-3) were centered at the sky frequencies of $\sim$683.6, 685.4, 679.9, and 681.8\,GHz, respectively, each with a bandwidth of about 2\,GHz. The data were collected thrice (two in Cycle-2, program ID: 2013.1.00524.S, PI: N. Lu, and one in Cycle-3, Program ID: 2015.1.00385.S, PI: N. Lu) in excellent weather conditions ($\textrm{precipitable water vapor}=0.24\,\textrm{mm,}\,0.39\,\textrm{mm and }0.40\,{\rm mm}$, respectively) but about 1 year apart from each successful observation. During the observations, the relatively extended configurations C34-5 and C36-4 , using up to 32, 37 and 42 12-m antennae (5, 2, and 0 out of which had problematic data; Table \ref{obslog}), were adopted. The total on-source integration time was $\sim$101 minutes. During the observations, the phase and gain variations were monitored using J2258-2758, J2258-2758 and J2151-3027, respectively. %The error in the flux calibration was estimated to be less than 10\% %(since the variation of calibrated source at 10% level was observed).

The data were reduced with CASA 4.6.0 (McMullin et al. 2007). The primary beam is $8^{\prime\prime}.8$, and the maximum recoverable scale is $\sim$$2^{\prime\prime}.1$. The calibrated datasets were combined and cleaned using the Briggs weightings with {\it robust=0.5}, and have nearly identical synthesized beams for the line and continuum emission, with the full width of half maximum (FWHM) of $\sim$$0^{\prime\prime}.15\times0^{\prime\prime}.13$, corresponding to physical scales of 37\,pc$\times$32\,pc, and a position angle (P.A.; north to east) of $-69^\circ$. The continuum was measured using data in SPWs 0-3 by excluding the line emission channels. For the \midco\ line emission, the cube was generated using the data in SPW-0, which encompasses the \midco\ emission at the systematic velocity (3422\,\kms; optical) with an effective bandpass of about 750\,\kms. The astrometric accuracy of these ALMA observations is better than $0^{\prime\prime}.07$, whereas the relative position accuracy is about $0.5\times$synthesized beam/signal-to-noise ratio{\footnote {\url{https://help.almascience.org/index.php?/Knowledgebase/Article/View/153/6/what-is-the-astrometric-position-accuracy-of-an-alma-observation}}}. Therefore, the relative position accuracies of the peak emission are about 5 and 6 mas, %(s/n=11.85 for line)
for our continuum and integrated line emission maps, respectively. 

Since there are large time gaps among our observations, and the flux calibrators for each observation are different, we also imaged each dataset separately to check quality and flux consistency. For each dataset, we found that the synthesized beams (area) are $2-3$ times larger than the aforementioned value, and the flux varies by an factor of $\sim$15\% (37\%) for the line (continuum) emission. Unless otherwise stated, flux measurements are based on the images after the primary beam correction, whereas all of the figures are produced using the results before the primary beam correction.

In order to increase the signal-to-noise ratio, we binned spectral cubes into channels with the width of $\delta v=14$\,\kms. The noise for these channel maps in \midco\ is in the range of $3.3-4.0$\,mJy\,beam$^{-1}$. For the continuum, the $1\sigma$ rms noise is about 0.4\,mJy\,beam$^{-1}$, and for the integrated \midco\ line emission map, which is integrated over the barycentric velocity range of $v=3214$$-$3522\,\kms\ in the datacube of 14\,\kms\ channel maps, it is 0.46\,Jy\,\kms\,beam$^{-1}$. All noise measurements were performed on the maps before the primary beam correction.

\section{Results and Discussion}
                                                                    
\subsection{\midco\ Line Emission}

\begin{figure*}[tb]
\centering
\includegraphics[width=0.8\textwidth,bb = 29 132 972 992]{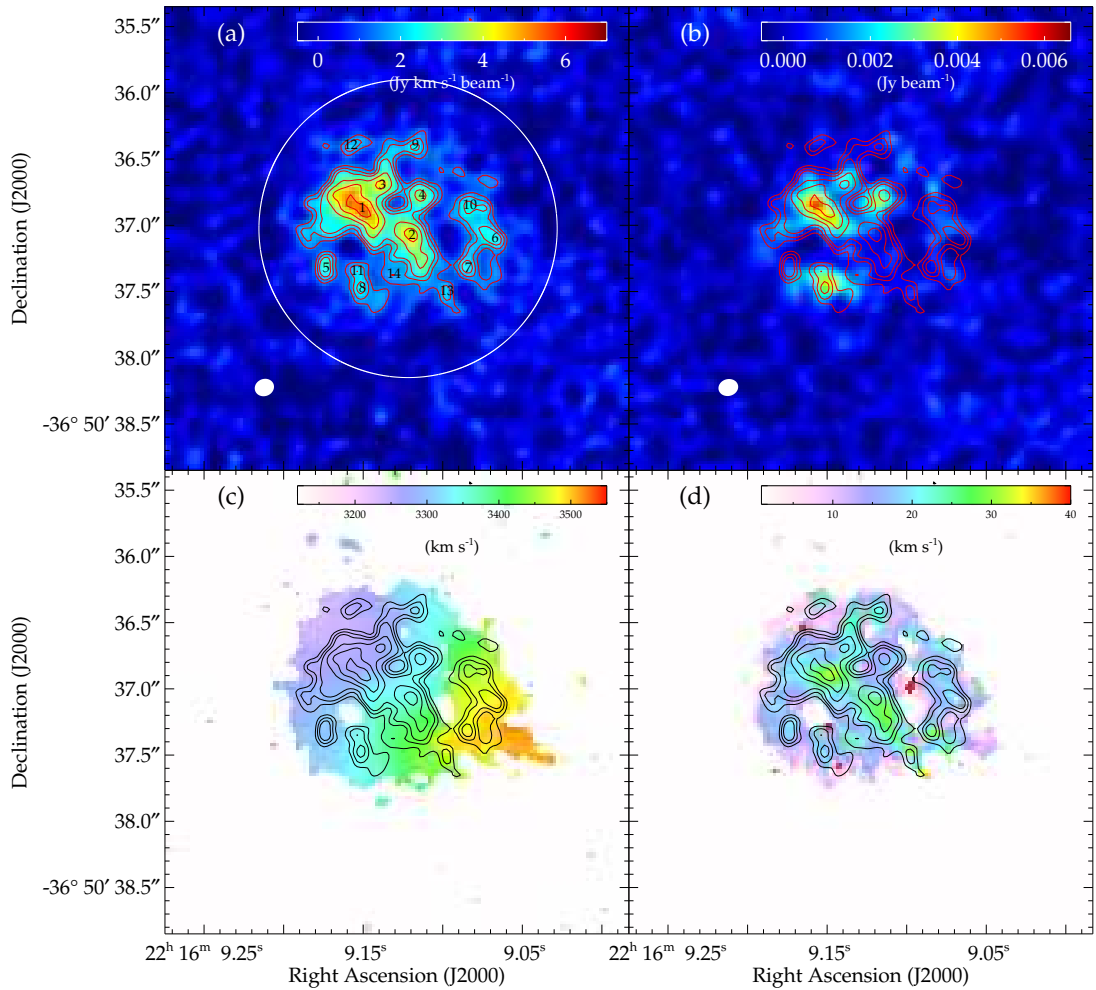}
\caption{\midco\ line emission contours of the integrated map superimposed on (a) the integrated \midco\ map; (b) the 434\,\mum\ continuum; (c) the velocity field map; and (d) the velocity dispersion map. The contour levels are [3, 4, 5, 7, 9, 11]\,$\times$\,$\sigma$ ($\sigma=0.46$\,Jy\,\kms\,beam$^{-1}$). The circle in (a) gives the region used to measure the total flux and extract the spectrum, and the (black) numbers are the clumps listed in Table \ref{clumpinfo}. The filled (white) ellipse in (a) and (b) show the beam shapes.}
\label{Figmom}
\end{figure*}

Figures \ref{Figmom}a-\ref{Figmom}d show the integrated \midco\ emission, the 434\,\mum\ continuum, the velocity field and velocity dispersion maps, respectively. All images are overlaid by the same contours of the integrated \midco\ line emission. The overall CO image looks clumpy, which may be due to the fact that (1) dense molecular gas is sparsely distributed in galaxies; and (2) our high resolution observation resolves out the largest-scale structures of $\gtrsim 2 \arcsec$ and excludes weak/extended morphologies from being revealed in our maps. %Furthermore, instead of discrete GMCs, we obtained a complicated filamentary web, which has also been observed in the Galactic star formation regions (e.g. Molinari et al. 2010). Our result suggests that GMCs in (U)LIRGs do not exist as a single super GMC, which has usually been used to classify (U)LIRGs as categorically different from normal galaxies. Indeed, if ISM is in filaments, then ULIRGs just have much higher-density filaments than normal nuclei. 
Furthermore, instead of discrete GMCs, we observed a complicated filamentary web, similar to what has been found in Galactic star formation regions (Molinari et al. 2010). %The high resolution CO~(3$-$2) map of Arp 220 also shows that the central kpc disc, which encompasses the two merging nuclei, has a filamentary structure (Sakamoto et al. 2008). This calls into question the bi-model scenario of the star formation law (Genzel et al. 2010), which argues that in normal galaxies star formation occurs in individual GMCs while in ULIRGs the density of GMCs is so high that they merged into a single cloud (or disc). Indeed, if the ISM is in filaments rather than in individual clouds, the ULIRGs may just have higher density filaments compared to those in Galactic star formation regions.
 
Using the clump-finding algorithm of Williams et al. (1994), which performs a blind search for clumps by contouring the map at different levels to identify peaks, we identified 14 clumps  and labelled them in Figure \ref{Figmom}a. The properties of these clumps, such as their positions, integrated fluxes, sizes (uncorrected for the beamsize), and velocity dispersions, are given in Table \ref{clumpinfo}. From the table we can see that only 3 of these 14 clumps are resolved by our observation. This might be due to the fact that the \midco\ emission practically traces warm and dense molecular cores which are significantly more compact than the angular resolution of the observations. In the Galactic center, the warm and dense molecular cores (density of $\sim$$10^4-10^5$\,cm$^{-3}$ and temperature of $\sim$60 to $>$100\,K) traced by the para-H$_2$CO emission are generally have sizes of $\lesssim$1\,pc (Ginsburg et al. 2016). 

The total flux of \midco\ measured within an circle of $r=1.125\arcsec$ (as plotted in Figure \ref{Figmom}a) from the integrated image is $160\pm16$\,Jy\,\kms, which is about 20\% of our SPIRE/FTS-measured flux of $802\pm52$\,Jy\,\kms\ (Lu et al. 2017), obtained with a much larger aperture of $\sim$33$^{\prime\prime}$. To further constrain the fraction recovered by our observation within the ALMA FOV, we adopted a simple method to estimate the total CO flux (and the dust continuum flux) within our FOV. Firstly, we assume that the \midco\ (or dust continuum at 434 \mum) to the 70 micron flux ratio ($R_{{\rm CO}/70\,\mu {\rm m}}$) is constant, according to Lu et al. (2014) and Liu et al. (2015). Secondly, we calculated $f_{70\,\mu {\rm m}}(\theta)$, the fractional 70 \mum\ flux within a Gaussian beam of FWHM $\theta$. For the \herschel\ FTS beam (FWHM$\sim$35\arcsec) at \midco, $f_{70\,\mu {\rm m}}(35\arcsec)$ is about 0.8 (Zhao et al. 2016a; Lu et al. 2017), and for the ALMA FOV $f_{70\,\mu {\rm m}}(9\arcsec)$ is $\sim$0.4. Therefore, our observation recovers 40\% ($\equiv 0.8/0.4 \times 20\%$) of the total CO flux within the ALMA FOV. However, this missing flux is a very crude estimation since $R_{{\rm CO}/70\,\mu {\rm m}}$ may vary at physical scales of $\lesssim$kpc (e.g., Liu et al. 2015).

Our result indicates that the majority of the \midco\ emission in IC 5179 is missed by our ALMA observation, though our previous studies show that in nearby LIRGs the \midco\ emission is highly concentrated in regions with a scale of half kpc (Xu et al. 2014, 2015; Zhao et al. 2016b). This could be attributed to the following reasons: (1)
%for a larger beam with (0.864x0.860 arcsec), the total flux is 255 Jy km/s, 1.4 times the current value
 Substantial \midco\ emission exists beyond the ALMA $\sim$9\arcsec\ field of view; (2) The extended emission with a scale $>$2\arcsec.1 ($\sim$520\,pc) is resolved out by our observation (the size of the line/continuum region is very close to 2\arcsec.1, which might be indirect evidence); (3) Weak diffuse features are below our sensitivity; and (4) a combination of these three. Indeed, the measured flux is 1.4 times higher than the current value when a much larger beam ($0\arcsec.86\times0\arcsec.86$) is used. This coarser resolution was obtained by doing the clean weighted with $robust=2.0$ and $uv$-taper ($\mathrm{outertaper}=100 k \lambda$). We also note that the total flux derived from the sum of the aperture photometry of individual channels (centered on the emission features for each given channel) is $\sim$$180\pm18$ Jy\,\kms\, which $\sim$13\% higher than that measured on the integrated \midco\ image. This is because, by co-adding all channel maps, the integrated emission is affected more severely by the (negative) side-lobes of different segments of the emission. This might be a significant effect since some segments are separated by 2\arcsec\ (see the channel maps in \S3.3), the angular scale limit of our interferometer observations.

\begin{deluxetable*}{cccccccc}{t}
\centering
\tablecaption{Information of the clumps identified in our \midco\ image\label{clumpinfo}}
\tablewidth{0pt}
\tabletypesize{\scriptsize}
\tablehead{\colhead{\multirow{2}{*}{No.}} & \colhead{R.A. (J2000)}& \colhead{Decl. (J2000)}&\colhead{FWHMx} &\colhead{FWHMy} &\colhead{$f_{\rm peak}$} &\colhead{$f_{\rm total}$}&$\sigma_v$\\
&(hh:mm:ss)&(dd:mm:ss)& (\arcsec) &(\arcsec) & (Jy\,\kms) &(Jy\,\kms)&(\kms)\\
(1)&(2)&(3)&(4)&(5)&(6)&(7)&(8)
}
\startdata
1&22:16:09.151&$-$36:50:36.90&0.37 (91)&0.29 (70)&0.16&35.4&43.6\\
 2&22:16:09.122&$-$36:50:37.07&0.24 (58)&0.29 (70)&0.13&20.4&31.7\\
 3&22:16:09.140&$-$36:50:36.70&0.28 (68)&0.19 (47)&0.13&13.0&28.6\\
 4&22:16:09.115&$-$36:50:36.77&0.17 (43)&0.16 (39)&0.10& 7.2&21.2\\
 5&22:16:09.176&$-$36:50:37.32&0.09 (22)&0.14 (34)&0.09& 3.1&22.1\\
 6&22:16:09.070&$-$36:50:37.10&0.16 (38)&0.18 (44)&0.08& 5.5&22.1\\
 7&22:16:09.086&$-$36:50:37.32&0.20 (48)&0.18 (45)&0.08& 4.3&32.7\\
 8&22:16:09.153&$-$36:50:37.47&0.17 (42)&0.14 (34)&0.07& 3.9&26.7\\
 9&22:16:09.118&$-$36:50:36.40&0.22 (54)&0.11 (26)&0.07& 4.1&29.2\\
10&22:16:09.084&$-$36:50:36.85&0.17 (42)&0.15 (36)&0.07& 5.5&24.3\\
11&22:16:09.155&$-$36:50:37.35&0.11 (26)&0.08 (19)&0.07& 1.8&27.3\\
12&22:16:09.159&$-$36:50:36.40&0.13 (31)&0.08 (20)&0.06& 1.7&17.0\\
13&22:16:09.099&$-$36:50:37.50&0.16 (39)&0.24 (57)&0.06& 3.5&34.8\\
14&22:16:09.132&$-$36:50:37.37&0.18 (43)&0.18 (43)&0.05& 3.0&33.1
\enddata
\tablecomments{Column 1: clump number; Columns 2 and 3: right ascension and declination of the clump center; Columns 4 and 5: FWHM at $x$- and $y$-direction, respectively, calculated at the level of $3\sigma$ and not corrected for the beam size; in the parentheses we also list the corresponding values in units of pc. Column 6: peak flux; Column 7: integrated flux;  Column 8: velocity dispersion, obtained by fitting a Gaussian profile to the spectrum extracted within an elliptical aperture with radii of (FWHMx, FWHMy).}
\end{deluxetable*}

\begin{figure}[t!]
\centering
\includegraphics[width=0.48\textwidth,bb = 88 2 621 618]{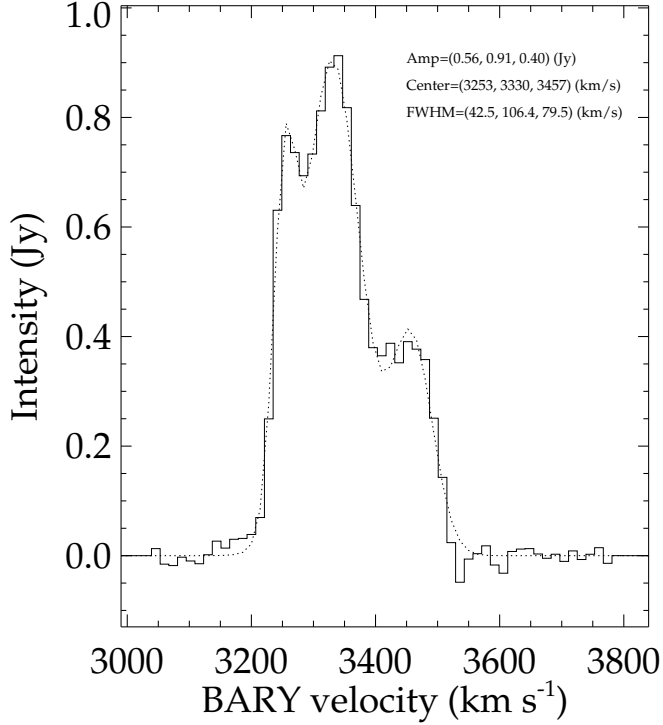}
\caption{Spatially integrated \midco\ line profile (after primary beam correction; solid line) taken within the white circle in Figure \ref{Figmom}. The dotted line gives the 3-Gaussian fitting result. The amplitude, central velocity and FWHM of each Gaussian function are labelled.}
\label{Figspec}
\end{figure}

Figure \ref{Figspec} shows the integrated CO spectrum for the region within the circle in Figure \ref{Figmom}a. The spectrum has multiple peaks, which is consistent with low-$J$ ($J=1,\,2$) CO observations (e.g., Mirabel et al. 1990; Garay et al. 1993). From the figure we can see that each component can be well fitted by a Gaussian function. The central velocity of the strongest peak in our spectrum is about 3330\,\kms. The flux-weighted mean, a measure of the systematic velocity of the kinematic center of the warm molecular gas disk, is 3345\,\kms. 
 
\begin{figure}[t!]
\centering
\includegraphics[width=0.47\textwidth,bb = 10 38 418 377]{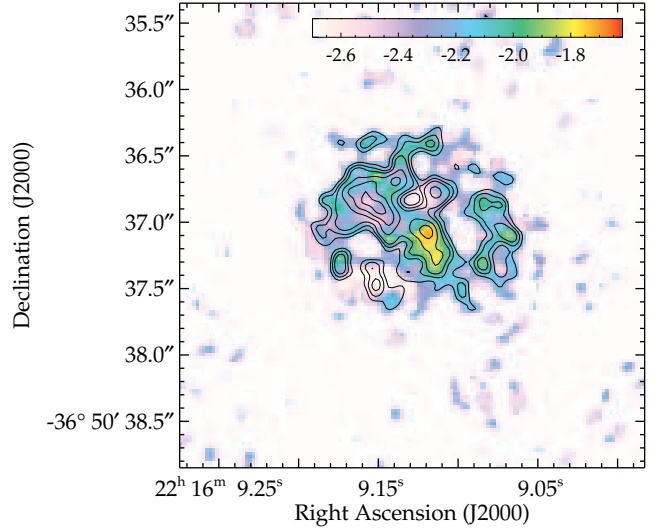}
\caption{Spatial variation of the \midco-to-dust continuum luminosity ratio (logarithm scale). We have used $\nu f_\nu({\rm cont})$ to calculate the continuum luminosity, where $f_\nu({\rm cont})$ is the flux of the continuum. The contours are the same as those in Figure \ref{Figmom}. See the text for more details.}
\label{Figco2dust}
\end{figure}

\subsection{Continuum Emission at 434\,\mum}
Our previous studies of the dust emission at $\sim$434\,\mum\ in NGC 34, NGC 1614 (Xu et al. 2014, 2015) and NGC 7130 (Zhao et al. 2016b) have shown that dust heating and gas heating in the warm dense gas cores are strongly coupled at large scales (e.g., $>$100 pc). As shown in Figure \ref{Figmom}b, the continuum generally correlates spatially with the \midco\ line emission as well in IC 5179. However, similar to what found in Arp 220 (Rangwala et al. 2015) and NGC 7130 (Zhao et al. 2016b), there also exist offsets ($0\arcsec.1-0\arcsec.3$, or 25$-$100 pc for different objects) between local peaks in the line and continuum emission. Furthermore, there is weak/no detectable dust emission in about a half of the CO-emission region (southwest). 

To further investigate the difference between the spatial distribution of the line and continuum emission, we plotted the \midco-to-continuum luminosity ratio (logarithm scale) in Figure \ref{Figco2dust}. Here we have adopted $\nu f_\nu ({\rm cont})$ as the monochromatic luminosity of the continuum emission. For pixels with fluxes less than 2$\sigma$, we used the 2$\sigma$ flux to calculate the ratio. From Figure \ref{Figco2dust} we can see that the CO-to-dust luminosity ratio varies by a factor of $>$10. %The highest ratios exist in the regions where there is no detection of the dust emission, whereas the lowest ratios occur at the emission peaks of the dust.

As discussed in detail in Zhao et al. (2016b), the displacement between the dust and line emission peaks, and the variation of the CO-to-dust luminosity ratio in NGC 7130 can be understood if they are heated by different mechanisms, e.g, the dust is heated by the UV radiation from young massive stars, whereas (part of) the CO-emission gas is mainly heated by supernova-driven shocks (see, e.g., Rosenberg et al. 2015). This scenario generally agrees with the observed results in IC 5179 as well. For example, the gas in the southwest part has weaker corresponding Pa-$\alpha$ emission (see \S3.4), and higher velocity dispersion, indicating a weaker radiation heating by the UV radiation field and a stronger mechanic heating by turbulence/shocks.

Our ALMA-detected flux density of the continuum at 434 \mum\ is $f_{434\,\mu{\rm m,\,ALMA}}=71\pm7$ mJy. 
%the total flux using the larger beam (see above) is 164 mJy, 2.3 times the current value
By interpolating the \herschel\ SPIRE photometric measurements at 350 and 500\,\mum, we estimated that the total flux density at 434 \mum\ is 3.01 Jy, indicating that only $\sim$2.4\% (5\% if the scaled value of the total flux adopted; see \S3.1) of the total dust emission is detected by our ALMA observation. This fraction is about an order of magnitude lower than the interferometer-to-single-dish flux ratio of the \midco\ line emission, suggesting that the dust should be  substantially more extended than the warm dense gas in IC 5179. %This is also supported by the fact that the continuum flux would increase to 2.3 times the current value once measured in the image with the larger beam ($0\arcsec.86\times0\arcsec.86$) mentioned \S3.1.

\subsection{Kinematics of the Nuclear Region}

\begin{figure*}[tbhp]
\centering
\includegraphics[width=0.95\textwidth,bb = 11 18 636 839]{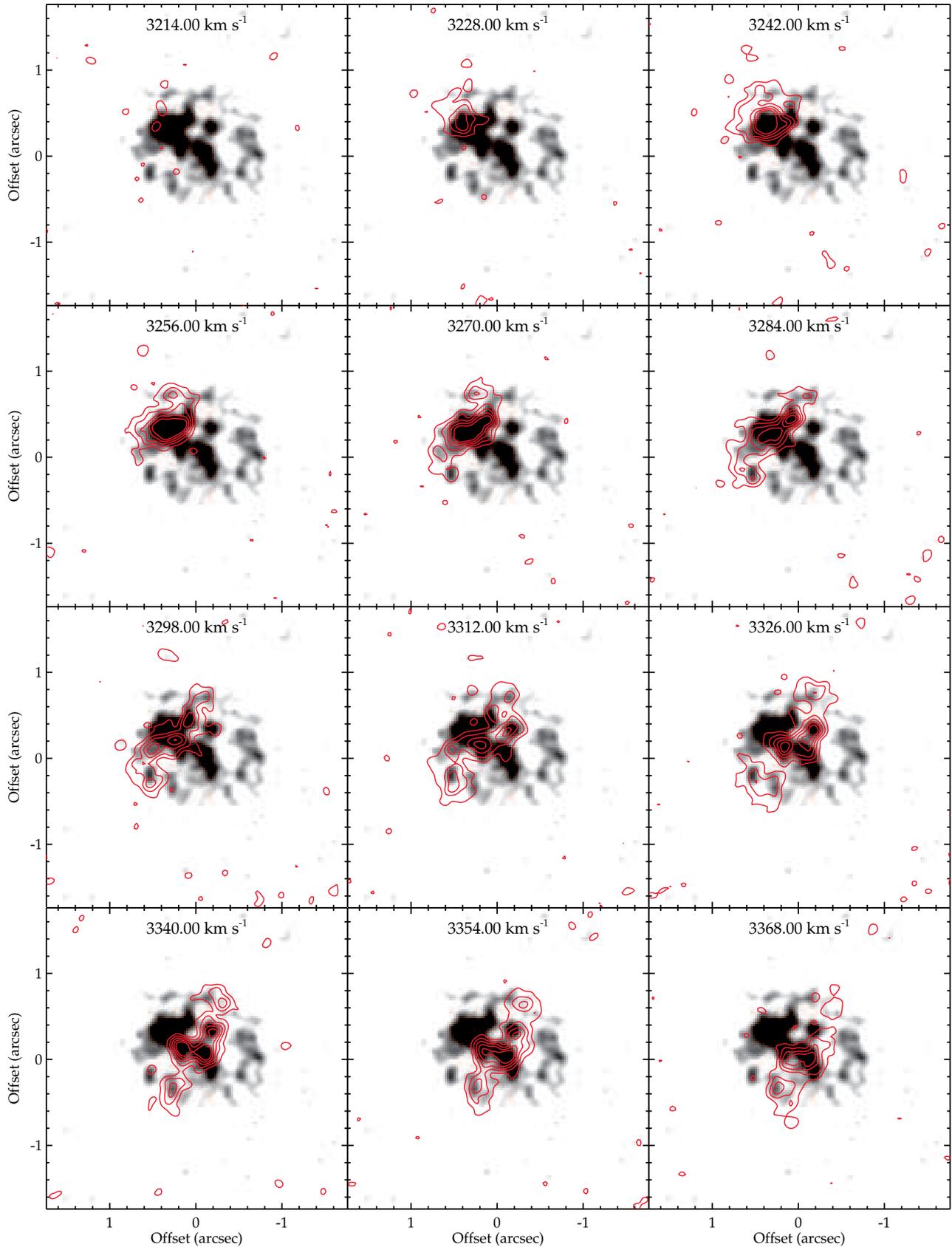}
\caption{\midco\ line emission contours of the channel maps (velocity channel width $\delta v=14$\,\kms), overlaid on the integrated emission image. The contour levels are [3, 6, 9, 12, 15]\,$\times$\,$\sigma\,(\sigma =3.3-4\,{\rm mJy\,beam}^{-1})$. In each channel, the central barycentric (radio) velocity is labeled.}
\label{Figchannel}
\end{figure*}

\setcounter{figure}{3}
\begin{figure*}[tbhp]
\centering
\includegraphics[width=0.95\textwidth,bb = 11 18 636 839]{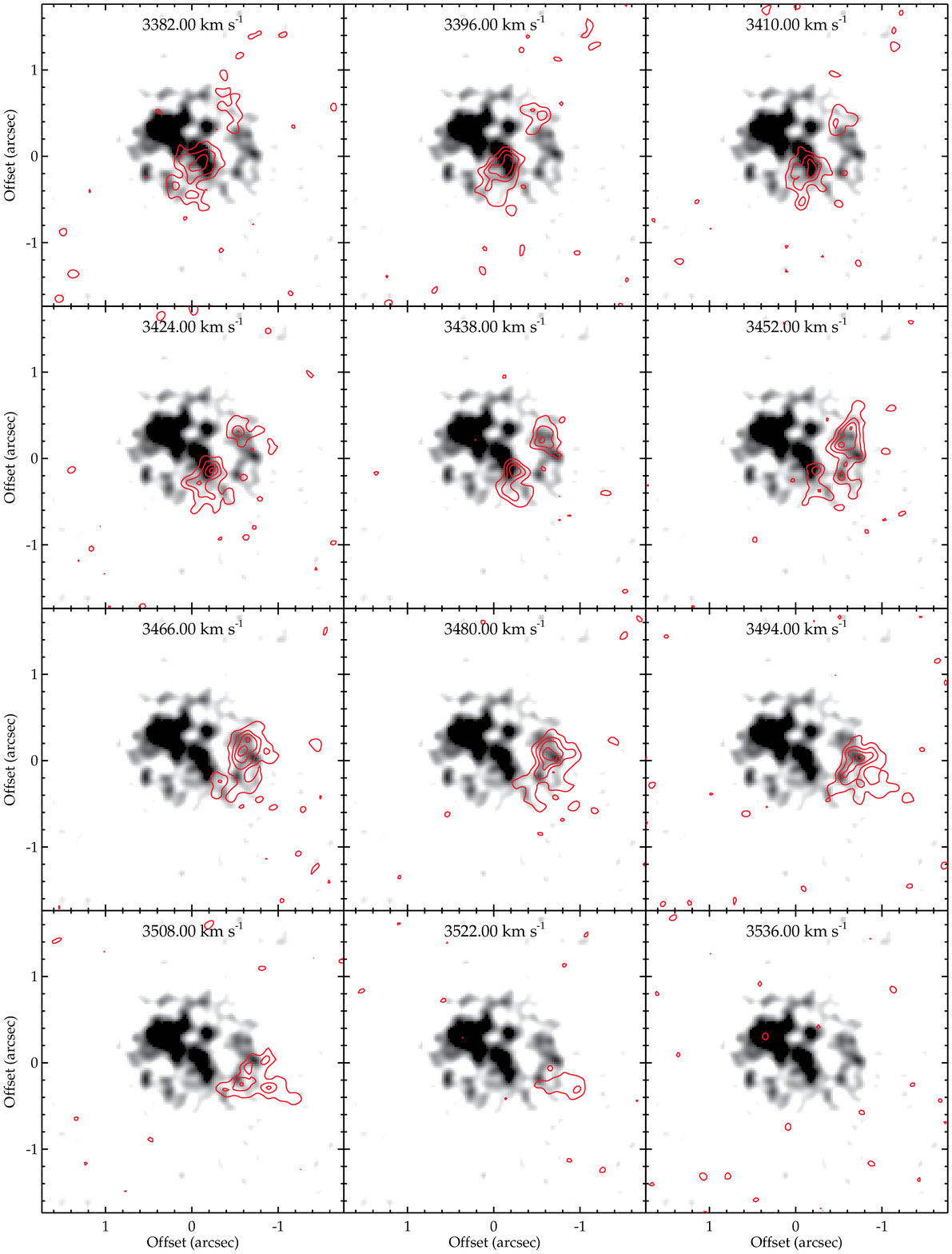}
\caption{--- Continued.}
\label{Figchannel}
\end{figure*}

\begin{figure}[t]
\centering
\includegraphics[width=0.47\textwidth,bb = 58 52 737 463]{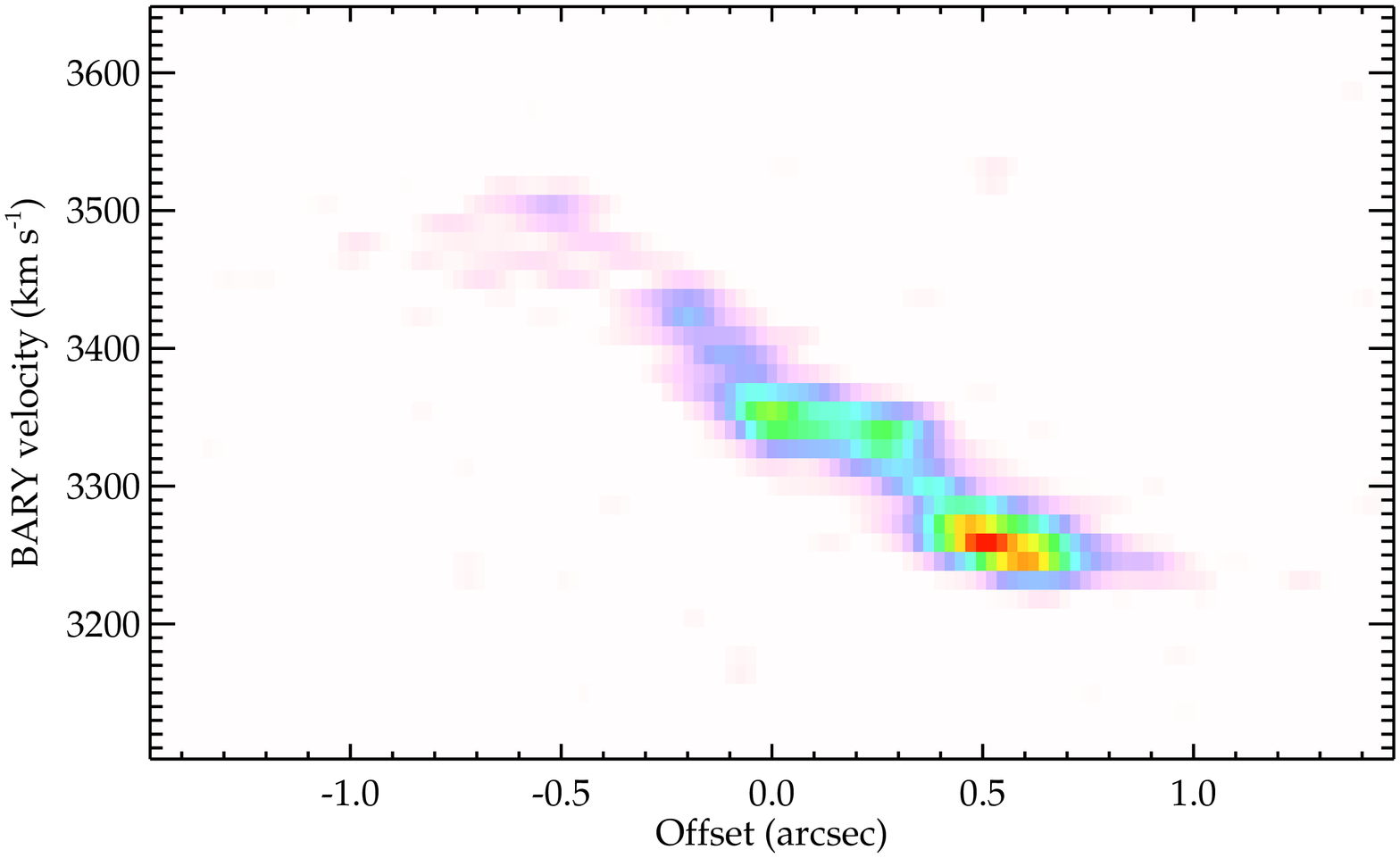}
\caption{P-V diagram extracted along a slice with the width of 0\arcsec.175 and P.A. of $\sim$55$^\circ$.}
\label{Figpv}
\end{figure}

The channel maps are displayed in Figure \ref{Figchannel}, where channels are overlaid on the integrated line emission. The spatial-velocity structure is demonstrated clearly in these plots. The velocity field (first moment) map (Figure \ref{Figmom}c) has the typical ``spider-diagram" shape, and shows a clear velocity gradient along the southwest-northeast direction, which  characterizes a rotating disk. From the position-velocity (P-V) diagram (Figure \ref{Figpv}), which was extracted using a slice with the width of 0\arcsec.175 (7 pixels) and P.A. of $\sim$55$^\circ$ (similar to the optical P.A.), the velocity amplitude (uncorrected for the inclination) of this rotation is $\sim$140\,\kms. To have a more detailed view, we further quantify the kinemetric features using the code {\it kinemetry}{\footnote{{\url http://davor.krajnovic.org/idl/}}}, which does surface photometry to the first  moment map by determining the best-fitting ellipse, and extracts information such as mean velocity, rotation curve, kinematic P.A., axial ratio, deviation from simple rotation, etc (see Krajnovi\'{c} et al. 2006 for more details).

Figure \ref{Figkin}a presents the kinematic P.A. as a function of radius $R$,  where the center (0.0 arcsec) was set to be (22$^{\rm h}$16$^{\rm m}$09$^{\rm s}$.132, $-36\degr50^\prime37\arcsec.00$). We can see that P.A. shows large variation at small radii (i.e. $R<0.4^{\prime\prime}$), indicating a twists in the isovelocity contours, and becomes stable at larger radii. The median value of the kinematic P.A. is $\sim$55$^\circ$, very similar to the one measured in the optical image (57$^\circ$; Lehnert \& Heckman 1995). Figure \ref{Figkin}b displays the variation of the axis ratio ($q$) along the radial direction. $q$  has similar behavior to P.A., and has a median value of 0.59, suggesting an inclination $i=\arccos q \sim 54^\circ$. The rotation curve (uncorrected for the inclination) shown in Figure \ref{Figkin}c illustrates that it arises fast within the central $0\arcsec.6$, and then flattens. The maximum $v_{\rm rot}$ is similar to that estimated from the P-V diagram (Figure \ref{Figpv}). In Figure \ref{Figkin}d we also plot the overall magnitude of the higher-order term ($k_5$), which is a kinemetric analogous of the photometric term that describes the deviation of isophote shape from an ellipse, and is sensitive to the existence of separate kinematic components on the velocity map (Krajnovi\'{c} et al. 2006). From the figure we see that the relative magnitude of $k_5$ can be as large as $\gtrsim$10\% of $v_{\rm rot}$ ($\equiv k_1$) when $R\gtrsim1\arcsec.1$, indicating that there might exist some weak non-circular motions (possibly due to infollows/outflows) in the gas disk.

Using long-slit and integral field spectroscopic data of the H$\alpha$ emission, Lehnert \& Heckman (1995) and Bellocchi et al. (2013) derived a maximum rotation speed of 194 and 186\,\kms, respectively. Comparing with our result, we can see that the rise of the velocity is almost confined to the central half kpc. This result can be confirmed by a close inspection of the first moment map in Bellocchi et al. (2013), though their data have a much lower spatial resolution.  

In the second moment map shown in Figure \ref{Figmom}d, the velocity dispersion ($\sigma_v$) in most regions is in the range of $10-30$\,\kms. However, regions with stronger emission  (i.e., $>$$3\sigma$; see the contours) in the zero moment map seem to have higher $\sigma_v$, with values of $20-35$\,kms. From the P-V diagram, the velocity gradient due to the rotation is estimated to be $dV/dr$$\sim$0.9\,\kms\,pc$^{-1}$, corresponding to a line widening of about 9\,\kms\ within individual beams (linear size of $\sim$34\,pc). This is consistent with the smallest $\sigma_v$ shown in the velocity dispersion map (Figure \ref{Figmom}d).  

\subsection{Comparison with Pa-$\alpha$ Observations}
\begin{figure}[t]
\centering
\includegraphics[width=0.47\textwidth,bb = 35 30 483 681]{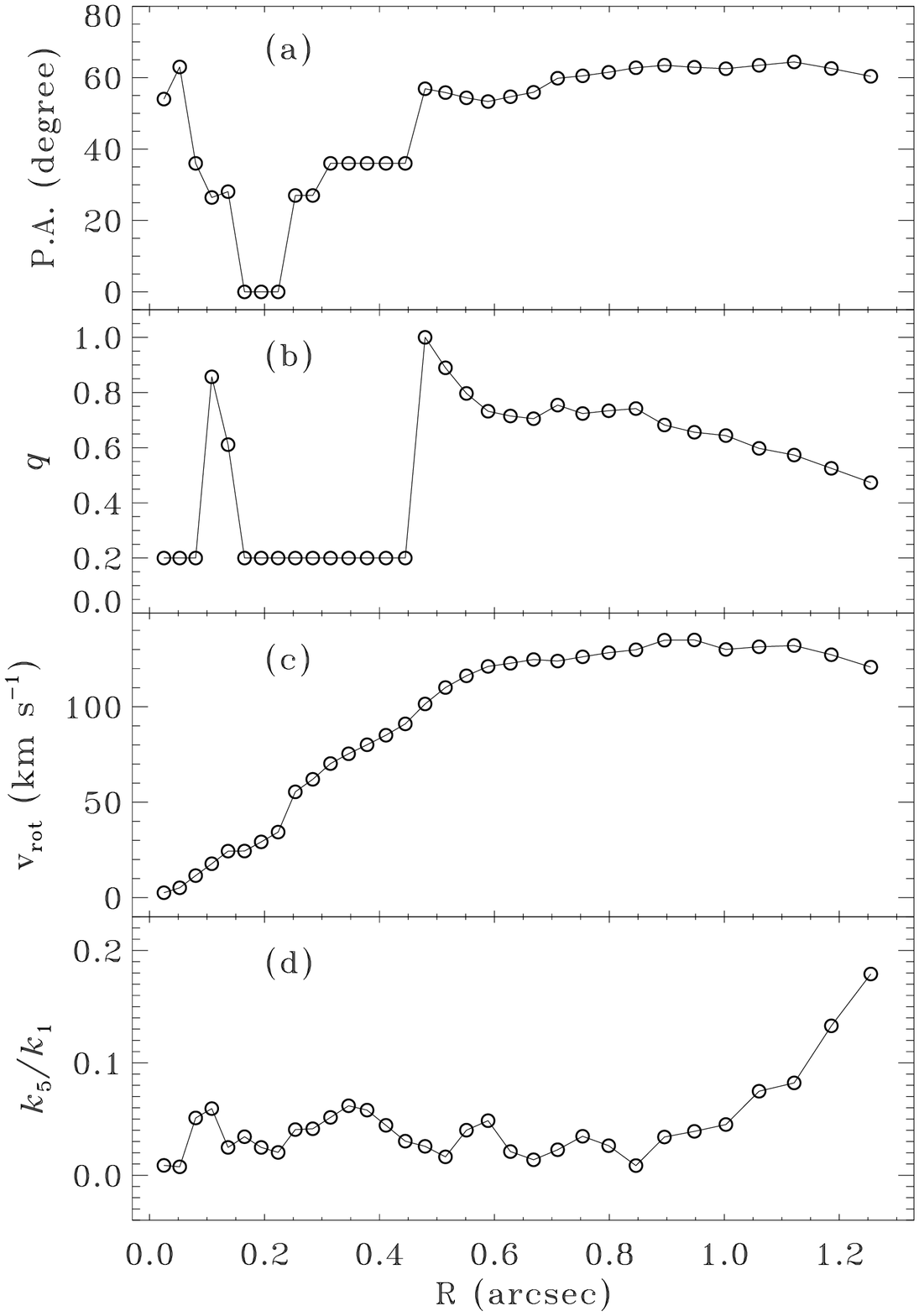}
\caption{Kinemetric coefficients of the \midco\ velocity map. (a) kinematic P.A.; (b) flattening $q$, e.g. the axis ratio; (c) projected rotation speed; and (d) magnitude of non-circular motions, i.e., the deviation from simple rotation.}
\label{Figkin}
\end{figure}

\begin{figure*}[t]
\centering
\includegraphics[width=0.80\textwidth,bb = 3 175 746 539]{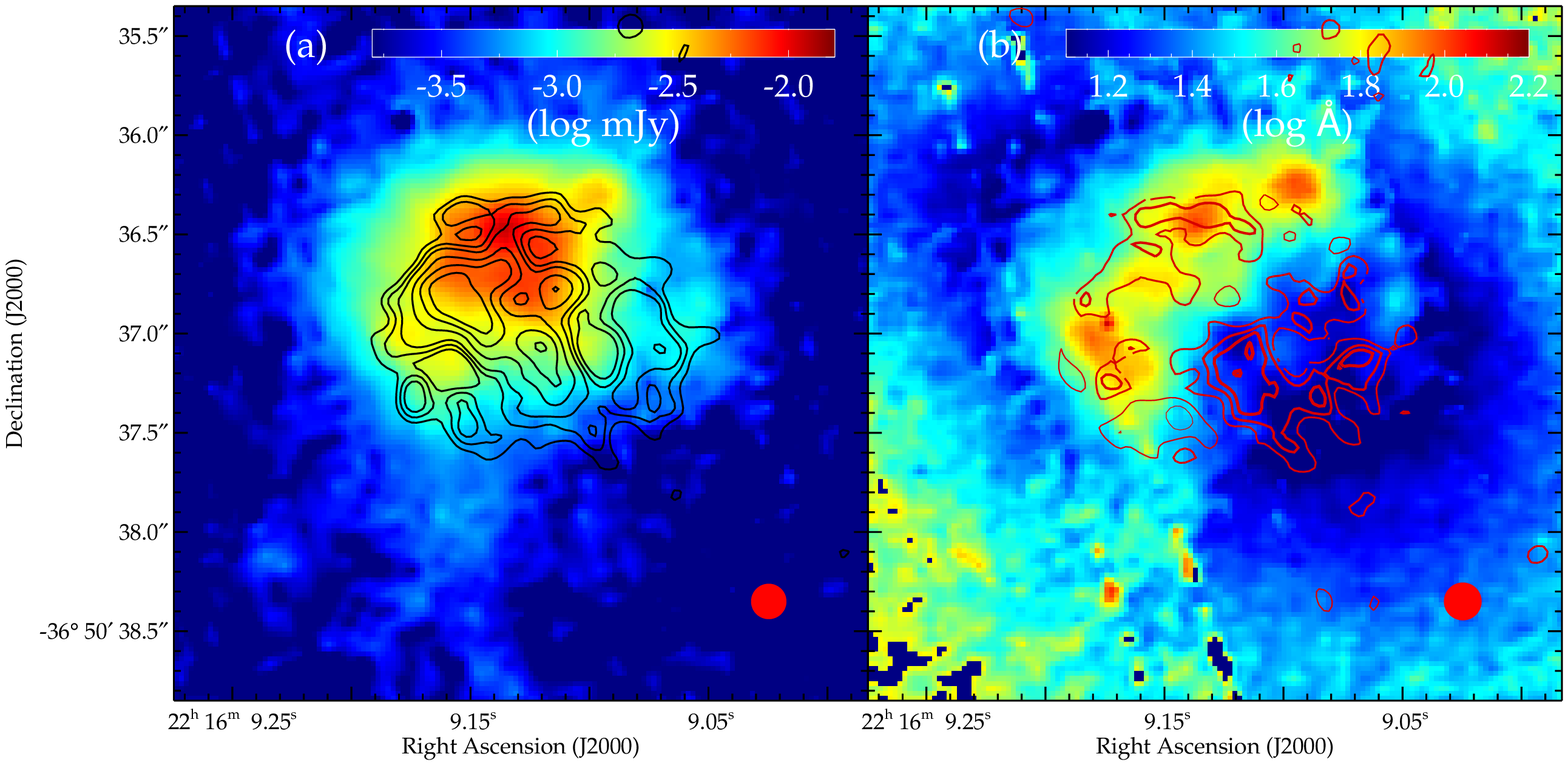}
\caption{(a) Integrated \midco\ line emission contours overlaid on the HST Pa-$\alpha$ image. The contour levels are $[3, 4, 5, 7,9, 11]\times \sigma$ with $\sigma=0.52$\,mJy\,beam$^{-1}$. (b) CO-to-continuum ratio contours overlaid on the Pa-$\alpha$ equivalent width (EW; in units of \AA) image. The contour levels are $[-2.75, -2.55, -2.25, -2.1, -1.95, -1.75]$ in logarithm, as indicated by their thickness (from thin to thick). The filled (red) circle in the bottom left of each panel illustrates the resolution of the image, where we have matched the resolutions (see the text for more details).}
\label{Figcomp}
\end{figure*}
In Figure \ref{Figcomp}a we plot the high-resolution ($0\arcsec.18$) {\it HST} Pa-$\alpha$ (Alonso-Herrero et al. 2006; Di\'az-Santos et al. 2008) image. The Pa-$\alpha$ map was obtained from a set of {\it HST} NIR narrow-band images (Di\'az-Santos et al. 2008). We have matched their resolutions by adopting $robust=2.0$ when cleaning the calibrated images of the CO emission. As shown in the figure, the \midco\ line emission roughly coincides spatially with the Pa-$\alpha$ emission, though their corresponding peaks do not agree. However, the Pa-$\alpha$ emission seems to vary smoothly with radius, with no clumpy morphology as seen in the \midco\ image. Furthermore, the southwest part of the CO-emission region has much weaker Pa-$\alpha$ emission. This should not be due to the extinction since the dust continuum is much weaker (see Figure \ref{Figmom}b) and the CO-to-continuum ratio is higher (Figures \ref{Figco2dust} and \ref{Figcomp}b) in this region.

Figure \ref{Figcomp}b shows the Pa-$\alpha$ equivalent width (EW; free from the extinction effect) image overlaid by CO-to-dust continuum ratio contours, where we have also adopted 2$\sigma$ as upper limits. As illustrated in D\'{i}az-Santos et al. (2008), the Pa-$\alpha$ EW is a useful age indicator of young stellar populations. It is interesting to find that the southwest region, where the CO-to-dust ratio is generally higher, shows about one order of magnitude smaller Pa-$\alpha$ EW than the northeast region, suggesting an older stellar population. This result, combining with the increasing non-circular motions (Figure \ref{Figkin}d), provides indirect evidence that the gas in the southwest region could be mainly excited by mechanical heating  rather than UV photons from young massive stars.

\section{Summary}
In this paper we present our ALMA observations of the \midco\ line emission and the continuum emission at $\sim$434\,\mum\ in the nucleus of the nearby LIRG IC 5179, with a physical resolution of about $37\,{\rm pc}\times32\,{\rm pc}$. Our main results are:
\begin{enumerate}
\item
The \midco\ emission displays a filamentary structure where many dense cores can be located, and shows a rotating disk-like velocity field. Moreover, its rotation curve reaches 90\% of the maximum value within the central $\sim$150\,pc.

\item
Globally the gas and dust are spatially correlated. However, their peaks do not always coincide with each other, and the CO-to-continuum ratio varies by at least an order of magnitude. This observed small scale variation can be attributed to the variation in the excitation mechanism of the CO line, as evidenced also by the values of the Pa-$\alpha$ EW.

\item
Within the nuclear region of $R\sim$300 pc and with a physical resolution of $\sim$34\,pc, our ALMA observations detected CO flux (dust flux density) is $180\pm18$\,Jy\kms\ ($71\pm7$\,mJy), only accounting for 22\% (2.4\%) of the total value derived from the \herschel/SPIRE observations.

\end{enumerate}

\begin{acknowledgements}
 We thank the anonymous referee for useful comments/suggestions that improve the paper. This work is supported in part by the NSFC grant NOs. 11673057, 11420101002, 11673028, 11643003 and 11503013. T.D-S. acknowledges support from ALMA-CONICYT project 31130005 and FONDECYT regular project 1151239. This paper makes use of the following ALMA data: ADS/JAO.ALMA\#2013.1.00524.S and \#2015.1.00385.S. ALMA is a partnership of ESO (representing its member states), NSF (USA), and NINS (Japan), together with NRC (Canada) and NSC and ASIAA (Taiwan), in cooperation with the Republic of Chile. The Joint ALMA Observatory is operated by ESO, AUI/NRAO, and NAOJ. This research has made use of the NASA/IPAC Extragalactic Database (NED), which is operated by the Jet Propulsion Laboratory, California Institute of Technology, under contract with the National Aeronautics and Space Administration.
\end{acknowledgements}

%\end{CJK*}
\end{document}